\documentclass[a4paper,fleqn,usenatbib]{mnras}

\usepackage[T1]{fontenc}
\usepackage{ae,aecompl}

\usepackage{graphicx}	
\usepackage{amsmath}	
\usepackage{amssymb}	

\newcommand {\magg}{\hphantom{$3$}}  

\newcommand{\nup}{$\nu_{\rm peak}^S$}

\newcommand{\en}{E_{\nu}}

\title[Extreme blazars as counterparts of IceCube neutrinos]{Extreme blazars as counterparts of IceCube astrophysical neutrinos}

 \author[P. Padovani et al.]{P. Padovani$^{1,2}$\thanks{E-mail:
ppadovan@eso.org}, E. Resconi$^3$, P. Giommi$^{4,5}$, 
 B. Arsioli$^{4,5,6}$, Y. L. Chang$^{4,6}$\\
$^{1}$European Southern Observatory, Karl-Schwarzschild-Str. 2,
D-85748 Garching bei M\"unchen, Germany\\
$^{2}$Associated to INAF - Osservatorio Astronomico di Roma, via Frascati 33,
I-00040
Monteporzio Catone, Italy\\
$^{3}$Technische Universit{\"a}t M{\"u}nchen, Physik-Department, James-Frank-Str. 1, 
D-85748 Garching bei M{\"u}nchen, Germany\\
$^{4}$ASI Science Data Center, via del Politecnico s.n.c., I-00133 Roma Italy \\
$^{5}$ICRANet-Rio, CBPF, Rua Dr. Xavier Sigaud 150, 22290-180 Rio de Janeiro, Brazil\\
$^{6}$Sapienza Universit\`a di Roma, ICRA, Dipartimento di Fisica, Piazzale Aldo Moro 5, 
I-00185 Roma, Italy}

\date{Accepted 2016 January 25. Received 2016 January 22; in original form 2015 December 18}

\pubyear{2015}

\begin{document}
\label{firstpage}
\pagerange{\pageref{firstpage}--\pageref{lastpage}}
\maketitle

\begin{abstract}

We explore the correlation of $\gamma$-ray emitting blazars with IceCube neutrinos
by using three very recently completed, and independently built, catalogues
and the latest neutrino lists. We introduce a new observable, namely the
number of neutrino events with at least one $\gamma$-ray counterpart,
$N_{\nu}$. In all three catalogues we consistently observe a positive
fluctuation of $N_{\nu}$ with respect to the mean random expectation at a
significance level of $0.4 - 1.3$ per cent. This applies only to extreme blazars,
namely strong, very high energy $\gamma$-ray sources of the high energy
peaked type, and implies a model-independent fraction of the current
IceCube signal $\sim 10 - 20$ per cent. An investigation of the hybrid photon --
neutrino spectral energy distributions of the most likely candidates
reveals a set of $\approx 5$ such sources, which could be linked to the
corresponding IceCube neutrinos. Other types of blazars, when testable,
give null correlation results. Although we could not perform a similar correlation study 
for Galactic sources, we have also identified two (further) strong Galactic $\gamma$-ray
sources as most probable counterparts of IceCube neutrinos through their hybrid 
spectral energy distributions. We have reasons to believe that our blazar results are not
constrained by the $\gamma$-ray samples but by the neutrino statistics,
which means that the detection of more astrophysical neutrinos could turn
this first hint into a discovery. 
\end{abstract}




\section{Introduction}\label{sec:Introduction}
The IceCube South Pole Neutrino
Observatory\footnote{http://icecube.wisc.edu} has recently reported the first
observations of high-energy astrophysical neutrinos\footnote{In this paper
  neutrino means both neutrino and antineutrino.}
\citep{2013PhRvL.111b1103A,ICECube13,ICECube14}. More recently, it has
confirmed and strengthened these observations by publishing a sample of 54
starting events collected over about four years and with a deposited
energy up to 2 PeV \citep{ICECube15_1}. These events are coming from the
entire sky and consist of neutrinos of all flavours which interact inside
the instrumented volume. The neutrino interaction vertex dominates the
signature of these events, the majority of which are shower-like.  The
complementary sample of through-going charged current $\nu_\mu$ from the
northern sky has been also studied over a period of two \citep{Aartsen2015}
and four years \citep{ICECube15_2} showing that the spectrum is inconsistent
with the hypothesis of purely terrestrial origin at 3.7$\sigma$ and 4.3$\sigma$ 
level respectively. These track-like events confirm the general picture
of a diffuse isotropic neutrino background although their energy spectrum
$E^{-\gamma}$ is harder ($\gamma = 1.91\pm0.20$) with respect to the all
sky one obtained from the starting events sample ($\gamma = 2.58\pm0.25$),
suggesting a mixed origin of the signal observed by IceCube.

Many diverse scenarios for the astrophysical counterparts of IceCube
neutrinos have been put forward \citep[see, e.g.][for a comprehensive
  discussion]{Ahlers_2015} but none has so far been statistically
supported by the observational data described above. 
One of the candidate neutrino-emitting astronomical classes of sources is that of blazars.
These are Active Galactic Nuclei (AGN) hosting a jet oriented at
a small angle with respect to the line of sight with highly relativistic
particles moving in a magnetic field and emitting non-thermal radiation
\citep{UP95}. The two main blazar sub-classes, namely BL Lacertae objects (BL Lacs) and
flat-spectrum radio quasars (FSRQ), differ mostly in their optical spectra,
with the latter displaying strong, broad emission lines and the former instead
being characterised by optical spectra showing at most weak emission lines,
sometimes exhibiting absorption features, and in many cases being completely
featureless. The general idea that 
blazars could be sources of high-energy neutrinos dates back to long
before the detection of sub-PeV neutrinos and
has since been explored in a number of studies \citep[e.g.][and papers from our group, 
as detailed below]{mannheim95,
  halzen97, mueckeetal03,kis14,muraseinouedermer14,tav15}. 

The spectral energy distributions (SEDs) of blazars are
composed of two broad humps, a low-energy and a high-energy one. The peak
of the low-energy hump (\nup) can occur at widely different frequencies,
ranging from about $\sim 10^{12.5}$~Hz ($\sim 0.01$ eV) to $\sim
10^{18.5}$~Hz ($\sim13$ keV). The high-energy hump, which may extend up to
$\sim 10$ TeV, has a peak energy that ranges between $\sim 10^{20}$~Hz
($\sim 0.4$ MeV) to $\sim 10^{26}$~Hz ($\sim 0.4$
TeV) \citep{GiommiPlanck,Arsioli2015}. Based on the rest-frame value of
\nup, BL Lacs can be further divided into Low energy peaked (LBL) sources
(\nup~$<10^{14}$~Hz [$<$ 0.4 eV]), Intermediate ($10^{14}$~Hz$<$ \nup~$<
10^{15}$~Hz [0.4 eV $<$ \nup $<$ 4 eV)] and High (\nup~ $> 10^{15}$~Hz [$>$
  4 eV]) energy peaked (IBL and HBL) sources respectively \citep{padgio95}.

\cite{Pad_2014} (hereafter PR14), on the basis of a joint positional and energetic
diagnostic using very high energy (VHE)\footnote{We adopt here the
  definitions used in \cite{2004vhec.book.....A} for $\gamma$-ray
  astronomy: ``high energy'' (HE) or GeV astronomy spans the 30 MeV to
  30 GeV energy range while VHE or TeV astronomy refers to the 30 GeV
  to 30 TeV range.} lists and studying $\gamma$-ray SEDs, have
suggested a possible association between eight BL Lacs (all HBL) 
and seven neutrino events reported by the IceCube collaboration
in 2014 \citep{ICECube14}. Following up on this idea,
\cite{Petro_2015} have modelled the SEDs of six of these BL Lacs using
a one-zone leptohadronic model and mostly nearly simultaneous data.
The SEDs of the sources, although different in shape and flux, were
all well fitted by the model using reasonable parameter
values. Moreover, the model-predicted neutrino flux and energy for
these sources were of the same order of magnitude as those of the
IceCube neutrinos. In two cases, i.e. MKN 421 and H 1914$-$194, a
suggestively good agreement between the model predictions and the
neutrino fluxes was found.

Very recently, \cite{Pad_2015} have calculated the cumulative neutrino
emission from BL Lacs ``calibrated'' by fitting the spectral energy
distributions of the sources studied by \cite{Petro_2015} and their
(putative) neutrino spectra. Within the so-called {\it blazar simplified
  view} \citep{paper1,paper2,paper3,paper4} and by adding a hadronic
component for neutrino production, BL Lacs as a class were shown to be able
to explain the neutrino background seen by IceCube above $\sim 0.5$ PeV
while only contributing on average $\sim 10$ per cent at lower energies. However, some room
was left for individual BL Lacs to still make a contribution at the
$\approx 20$ per cent level to the IceCube low-energy events.

The hypothesis put forward by PR14 and \cite{Petro_2015}, if correct,
should materialise in an IceCube detection but this has not happened
yet. At present, in fact, IceCube has not identified any point sources 
and therefore its signal remains unresolved. The published upper
limits on blazars start to be in the ballpark of the scenario described above (PR14) although 
they do not rule it yet out \citep{ICECube15_3}.

Together with the larger neutrino samples recently provided by the
IceCube Collaboration, new and better catalogues of high energy
sources are now available, which overcome some of the limitations
pointed out in PR 14, like the lack of an all-sky
flux-limited TeV catalogue. The purpose of this paper is to study in a
more quantitative way the possible connection between the IceCube
astrophysical neutrinos and $\gamma$-ray emitting blazars. To this
aim, we have selected a priori 2FHL \citep{2FHL} and 2WHSP (Chang et
al. 2015, in preparation) as the best VHE catalogues, as detailed
below. The {\it Fermi} 3LAC catalogue \citep{Fermi3LAC} was also used
because of its size and all-sky coverage, although it reaches
$\gamma$-ray photons of lower energy. We note that the scanning
strategy and the intervals over which the connection between 
neutrinos and $\gamma$-ray sources was studied have also been fixed before any
test was carried out.

Section 2 describes the neutrino and $\gamma$-ray catalogues used in
this paper, while Section 3 discusses our statistical
analysis. Section 4 gives our results, while in Section 5 we
investigate the $\gamma$-ray counterparts and their SEDs. Section 6
summarises our conclusions. Appendix A deals with the 2FHL Galactic
sources.

\section{The catalogues}

\subsection{Neutrino lists}\label{sec:neutrino_list}

\begin{table*}
\caption{Selected list of high-energy neutrinos detected by IceCube.}
\begin{tabular}{@{}cclllrr}
IceCube ID & Dep. Energy & $\nu f_{\nu}$$^a$ & RA (2000) & Dec (2000) & Median angular error & $b_{\rm II}$ \\ 
                      & TeV  & $10^{-11}~$erg/cm$^{2}$/s&  &   & deg  &   deg \\
\hline
\magg3     &   78.7$^{+10.8}_{-8.7}$ & 1.4$^{+3.3}_{-1.2}$ & 08 31 36 & $-$31 12 00 & $\le$1.4 & $+5$  \\
\magg4     &  165$^{+20}_{-15}$ & 0.8$^{+1.9}_{-0.7}$ & 11 18 00 & $-$51 12 00 &  7.1  & $+9$  \\
\magg5     &   71.4$\pm$9.0 &  1.3$^{+3.0}_{-1.1}$ & 07 22 24 & $-$00 24  00 & $\le$1.2  & $+7$  \\
 \magg9     &   63.2$^{+7.1}_{-8.0}$ & 2.1$^{+4.7}_{-1.7}$ & 10 05 12 &  $+$33 36 00 & 16.5 &  $+54$ \\
10   &   97.2$^{+10.4}_{-12.4}$ & 1.2$^{+2.8}_{-1.0}$ & 00 20 00 & $-$29 24 00&  8.1& $-83$  \\
11 &   88.4$^{+12.5}_{-10.7}$ & 1.1$^{+2.5}_{-0.9}$ & 10 21 12 & $-$08 54 00 & 16.7 & $+$39  \\
12 &  104$\pm13.0$& 0.9$^{+2.1}_{-0.8}$&19 44 24 & $-$52 48  00 &  9.8 & $-29$ \\
13 &  253$^{+26}_{-22}$&  1.2$^{+2.7}_{-1.0}$ & 04 31 36 & $+$40 18 00 & $\le$1.2 & $-5$ \\
14 & 1041$^{+132}_{-144}$& 1.1$^{+2.6}_{-0.9}$ & 17 42 24& $-$27 54  00&13.2 & $+1$ \\
17 &  200$\pm$27 & 1.2$^{+2.9}_{-1.0}$& 16 29 36 & $+$14 30  00  & 11.6 & $+38$ \\
19 &   71.5$^{+7.0}_{-7.2}$ & 1.3$^{+3.0}_{-1.1}$ & 05 07 36 & $-$59 42 00 &  9.7 & $-36$  \\
20 & 1141$^{+143}_{-133}$&  1.1$^{+2.6}_{-0.9}$ & 02 33 12 & $-$67 12 00& 10.7 & $-47$ \\
22 &  220$^{+21}_{-24}$&  0.7$^{+1.7}_{-0.6}$& 19 34 48 & $-$22 06 00& 12.1& $-19$ \\
23 &   82.2$^{+8.6}_{-8.4}$ & 1.5$^{+3.5}_{-1.3}$ & 13 54 48 & $-$13 12 00& $\le$1.9 & $+$47\\   
26 &  210$^{+29}_{-26}$ & 1.1$^{+2.6}_{-0.9}$ & 09 33 36 & $+$22 42 00  &11.8 & $+45$ \\
27 &   60.2$\pm5.6$ &  1.8$^{+4.0}_{-1.5}$ & 08 06 48 &  $-$12 36 00 &  6.6 & $+10$  \\
30 &  129$^{+14}_{-12}$& 0.8$^{+1.9}_{-0.7}$ & 06 52 48 & $-$82 42 00&   8.0 & $-27$ \\
33 &  385$^{+46}_{-49}$& 1.4$^{+3.2}_{-1.2}$  &19 30 00& $+$07 48 00& 13.5 & $-5$ \\
35 & 2004$^{+236}_{-262}$& 1.4$^{+3.3}_{-1.2}$ & 13 53 36 &$-$55 48 00 & 15.9 & $+6$ \\
38 &  201$\pm16$  & 1.2$^{+2.9}_{-1.0}$  & 06 13 12 & $+$14 00 00 & $\le$1.2 & $-2$ \\   
39 &  101$^{+13}_{-12}$ &  0.9$^{+2.0}_{-0.7}$ & 07 04 48 & $-$17 54 00 & 14.2 & $-$5 \\   
40 &  157$^{+16}_{-17}$ & 0.8$^{+1.8}_{-0.6}$ & 09 35 36 & $-$48 30 00 & 11.7 & $+$3 \\   
41 &   87.6$^{+8.4}_{-10.0}$ & 1.4$^{+3.2}_{-1.2}$ & 04 24 24 &$+$03 18 00 & 11.1 & $-$30 \\   
44 &   84.6$^{+7.4}_{-7.9}$  & 1.4$^{+3.1}_{-1.1}$  & 22 26 48 & $+$00 00 00 & $\le$1.2 & $-$46 \\   
45 &  430$^{+57}_{-49}$ & 0.9$^{+2.0}_{-0.7}$  & 14 36  00 & $-$86 18 00 &  $\le$1.2 & $-$24 \\   
46 &  158$^{+15}_{-17}$ & 0.8$^{+1.8}_{-0.7}$       & 10 02 00 & $-$22 24 00 &  7.6 & $+$26 \\   
47 &   74.3$^{+8.3}_{-7.2}$ & 1.6$^{+3.8}_{-1.4}$  & 13 57 36 & $+$67 24 00 &  $\le$1.2 & $+$48 \\   
48 &  105$^{+14}_{-10}$ & 0.9$^{+2.1}_{-0.8}$   & 14 12 24 & $-$33 12 00 &  8.1 & $+$27 \\   
51 &   66.2$^{+6.7}_{-6.1}$  &  2.2$^{+5.0}_{-1.8}$  &  05 54 24 & $+$54 00 00 &  6.5 & $+$14 \\   
52 &   158$^{+16}_{-18}$  & 0.8$^{+1.8}_{-0.7}$  & 16 51 12 & $-$54 00 00 &  7.8 & $-$6 \\   
    &    2600$\pm300$&        &  07 21 22 &  $+$11 28 48 &  0.27 & +12 \\
\hline
\multicolumn{7}{l}{\footnotesize $^a$ Fluxes in units of $10^{-8}$ GeV cm$^{-2}$ s$^{-1}$ can be obtained by
multiplying the numbers in this column by 0.614.}\\
\end{tabular}
\label{tab:ICE}
\end{table*}

This work is based on the IceCube high-energy starting events (HESE)
published by \cite{ICECube14} and \cite{ICECube15_1}, which cover the first
four years of data plus the $\nu_{\mu}$ selected from a large sample
of high-energy through-going muons \citep[see][and the IceCube online
  link\footnote{https://icecube.wisc.edu/science/data/HE\_NuMu\_diffuse}
  for the full list]{Aartsen2015}. Finally, we also included the very high energy
  (2.6 PeV deposited energy) event announced by the IceCollaboration 
  in July 2015 \citep{2015ATel.7856....1S}. 

Following PR14 we made the following two cuts to the HESE list:
1. neutrino energy $\en \ge 60$ TeV, to reduce the residual
atmospheric background contamination, which might still be produced by
mouns and atmospheric neutrinos and concentrates in the low-energy
part of spectrum \citep[see Fig. 2 in][]{ICECube14}; 2. median angular
error $\le 20^{\circ}$, to somewhat limit the number of possible
counterparts. The final list includes 30 HESE and 21\footnote{One of
  the $\nu_{\mu}$ events coincides with HESE ID 5 and was therefore
  discarded.} through-going $\nu_{\mu}$, for a total of 51 IceCube
events. The former, together with the 2.6 PeV event, are listed in
Tab. \ref{tab:ICE}, which gives the deposited energy of the neutrino,
the flux at the deposited energy in $\nu f_{\nu}$ units, the
coordinates, the median angular error in degrees, and the Galactic
latitude. For the through-going $\nu_{\mu}$, for which we refer the
reader to the online IceCube link, we assumed a median angular error
of $0.4^{\circ}$, as prescribed by the IceCube collaboration, apart from the 2.6 PeV
event, for which the median angular error is $0.27^{\circ}$  \citep{2015ATel.7856....1S}.

Neutrino fluxes have been derived as in PR14 but using a live time of
detection of 1,347 days \citep{ICECube15_1}. This means that the values for
the sources studied in PR14 are now smaller by a factor $1,347/998 =
1.363$. The derived fluxes are in the range $0.7 - 2.2 \times 10^{-11}$ erg
cm$^{-2}$ s$^{-1}$ (i.e., $0.4 - 1.3\times 10^{-8}$ GeV cm$^{-2}$ s$^{-1}$)
and errors are Poissonian for one event \citep{geh86}.

\subsection{$\gamma$-ray catalogues}

\subsubsection{{\it Fermi} 2FHL} 

The second catalogue of hard {\it Fermi}-Large Area Telescope (LAT)
sources \citep[2FHL:][]{2FHL} includes 360 sources and provides an
all-sky view of VHE sources at $E > 50$ GeV. We remind the reader that
1FHL \citep{1FHL}, the first {\it Fermi}-LAT catalogue of hard sources, had
a 10 GeV threshold, i.e. still not on the VHE side. 2FHL, instead,
bridges the gap between {\it Fermi}-LAT and ground based Cherenkov
telescopes. Given its all-sky nature we can use 2FHL  also
to select a sample of Galactic sources. We then defined two
subsamples: 1. the $|b_{\rm II}| \ge 10^{\circ}$ subsample, which
contains 257 objects, of which a very large fraction ($\sim 90$ per cent) are
blazars\footnote{These and the following numbers reflect our own
  classification of many of the unclassified 2FHL sources, using also
  2WHSP (see below), and are somewhat different from those given in
  the 2FHL paper.}. The remaining sources are mostly still
unclassified but very likely to be blazars; 2. the $|b_{\rm II}| <
10^{\circ}$ subsample, which contains 103 objects, of which a good
fraction ($\sim 41$ per cent) are still blazars. The remaining $59$ per cent is
composed of Galactic objects, that is supernova remnants (SNR) and
pulsar wind nebulae (PWN) ($\sim 33$ per cent), and unclassified sources
($\sim 26$ per cent), very likely to be Galactic as their VHE {\it Fermi}
spectrum is harder than that of extragalactic sources
\citep{2FHL}. The Galactic sources are discussed in Appendix
\ref{sec:Galactic}.

We further subdivided the $|b_{\rm II}| \ge 10^{\circ}$ subsample into
HBL (\nup $> 10^{15}$~Hz) and non HBL. The former sample contains 149
sources, all BL Lacs, while the latter, which is made up of 108
objects, contains mostly blazars of the IBL and LBL type ($\sim 69$ per cent,
including some FSRQs [$\sim 9$ per cent]), unclassified sources ($\sim
23$ per cent), and radio galaxies ($\sim 8$ per cent).



\subsubsection{2WHSP}\label{sec:2WHSP}

The 1WHSP catalogue \citep{Arsioli2015} provided a large area
($|b_{\rm II}| > 20^{\circ}$) catalogue of $\sim 1,000$ blazars and
blazar candidates selected to have \nup $> 10^{15}$ Hz and therefore
expected to radiate strongly in the HE and VHE bands. 1WHSP sources
were characterized by a ``figure of merit'' (FoM), which quantified
their potential detectability in the TeV band by the current
generation of Imaging Atmospheric Cherenkov telescopes.  This was
defined as the ratio between the synchrotron peak flux of a source and
that of the faintest blazar in the 1WHSP sample already detected in
the TeV band. 1WHSP sources are all BL Lacs, with a large fraction of
those with high FoM being known TeV sources (e.g., $36$ per cent of those
with FoM $\ge 1.2$) and the remaining ones thought to be within reach
of detection by current VHE instrumentation. Although technically not
a $\gamma$-ray catalogue, 1WHSP represented at the time the best way
to compensate for the lack of full sky coverage in the TeV band for
blazars. Moreover, $\sim 30$ per cent of the sources already had a {\it
  Fermi} 1FGL, 2FGL or 3FGL $\gamma$-ray counterpart.

Chang et al. (2015, in preparation) have updated the 1WHSP catalogue
and produced 2WHSP, which reaches down to $|b_{\rm II}| \ge
10^{\circ}$ and drops one of the previously adopted selection criteria
(the IR colour-colour cut) to increase completeness at low IR fluxes
and better include some HBL sources dominated in the optical and IR
bands by the light from the host giant elliptical galaxy. The 2WHSP
catalogue includes $\sim 1,700$ sources and therefore provides a $\sim
70$ per cent increase in size as compared to 1WHSP.  It reaches much lower
VHE fluxes, and it is almost seven times larger, than 2FHL.

The 2FHL and 2WHSP catalogues do not have the same composition in
terms of blazar types.  Of the 240 sources in the 2FHL catalogue that
are identified with a counterpart at other frequencies and are located
at $|b_{\rm II}| \ge 10^{\circ}$ only $\sim 71$ per cent are also in
2WHSP. The remaining objects are all blazars with \nup $< 10^{15}$ Hz,
which therefore cannot be included in the 2WHSP sample by
definition. On the other hand, $\sim 93$ per cent of the 2FHL HBL with $|b_{\rm II}| \ge 10^{\circ}$ 
are also part
of 2WHSP while $\sim 70$ per cent of the 2WHSP subsample with FoM $\ge 2$ are
2FHL HBL sources. We note that 2WHSP has an advantage over 2FHL, as it
not affected by extragalactic background light (EBL) absorption, since
the FoM is defined at \nup, while 2FHL is selected based on the flux
at $E > 50$ GeV. A relatively high redshift source, for example is
less likely to be in the 2FHL sample than in 2WHSP. 

\subsubsection{{\it Fermi} 3LAC}\label{sec:def_3LAC}

We also used the third catalogue of AGN detected above 100 MeV
by the {\it Fermi}-LAT \cite[3LAC:][]{Fermi3LAC}, more specifically the
``clean sample'' of 1,444 sources at $|b_{\rm II}| \ge 10^{\circ}$ and free
of the analysis issues, which affect some of the 3LAC detections. Basically
all objects ($\sim 98.8$ per cent) are blazars. We do not expect a neutrino signal
from this sample, however, at least as far as the full sample is concerned, 
based on the results of \cite{gluse_2015}, who
found no evidence of neutrino emission and a maximal contribution from {\it
  Fermi} 2LAC \citep{fermi2lac} blazars $\sim 20$ per cent. Moreover,
\cite{brown15}, using 70 months of {\it Fermi}-LAT observations, found no
evidence of $\gamma$-ray emission associated with IceCube's track-like
neutrino events. 

We further subdivided the 3LAC sample into an HBL (\nup $>
10^{15}$~Hz), an FSRQ, and an ``others" sample, which include 386,
415, and 645 sources respectively\footnote{These numbers sum up to
  1,446 (and not 1,444) because two FSRQ happen to be HBL}. The
``others'' sample is made up for the most part of BL Lacs and
``unclassified AGN'' of the IBL and LBL type ($\sim 97$ per cent), with the
remaining $\sim 3$ per cent including steep-spectrum radio quasars and radio
galaxies.

 
\section{The statistical analysis}\label{sec:stat}

To study the possible connection between the IceCube neutrinos and the
source catalogues, we have introduced the observable $N_{{\nu}}$ defined as
the number of neutrino events with at least one $\gamma$-ray counterpart
found within the individual median angular error. To evaluate this, we also took into account 
the case of IceCube neutrinos with $|b_{\rm II}| < 10^{\circ}$ but which could still be 
associated with a $|b_{\rm II}| >10^{\circ}$ $\gamma$-ray source given their large 
error radii. We do not only consider
the whole $\gamma$-ray catalogues but within a given catalogue we scan
versus flux, $N_{{\nu}}(f_\gamma)$ or, equivalently, versus FoM for
2WHSP, $N_{{\nu}}(FoM)$. If only the strongest sources are associated to
IceCube events such a scan will reveal a deviation from the randomised
cases.
The chance probability $P_i(N_{{\nu}}(f_\gamma,i))$, or equivalently
$P_i(N_{{\nu}}(FoM,i)$), to observe a certain $N_{{\nu}}$ for sources with
$f_\gamma \geq f_\gamma,i$ is determined on an ensemble of typically
$10^5$ randomised maps. Where needed, the sampling has been done over
$10^6$ randomised maps. To determine the random cases, we explored three
different procedures: 1. randomisation of the $\gamma$-ray sample
coordinates by drawing an equal number of positions homogeneously
distributed over the sky, making sure that only random sources with $b_{\rm
  II}$ values in the same range as the original $\gamma$-ray catalogue are
considered. This leaves untouched the IceCube positions, which are known to
be not uniformly distributed, but might lose any large scale structure
present in the $\gamma$-ray sample; 2.  to at least partially obviate 
this, we have also randomised only the $\gamma$-ray sample right
ascensions; 3. randomisation of the IceCube right ascensions, making sure
that only random sources that can be associated with the $\gamma$-ray
catalogue within the relevant error circles are considered (for example, a
$|b_{\rm II}| < 10^{\circ}$ IceCube random source can still be associated
with a 2WHSP object provided it has a large enough error radius). This
preserves any large scale structure present in the $\gamma$-ray sample plus
the IceCube declinations. However, the requirement above does not conserve
the total area sampled by the IceCube error circles, resulting in a biased
test statistics. As a result, we used procedures 1 and 2,
conservatively taking as our best estimate of the probability of random
association the {\it largest} of the two probabilities.
Since our three catalogues are somewhat overlapping, in particular at large fluxes 
and FoM (see Section \ref{sec:2WHSP}), the number of bins per scan is $< 10$, and the points are not independent, 
we do not correct the probabilities for the ``look elsewhere effect'' (which
would take into account the artificial p-value reduction due to the application of 
multiple tests). We have also fixed the catalogues studied at the beginning of this work as well
as the bins in flux and FoM used for the scans. 
\section{Results}

\subsection{{\it Fermi} 2FHL}

\begin{figure}
\includegraphics[height=9.0cm]{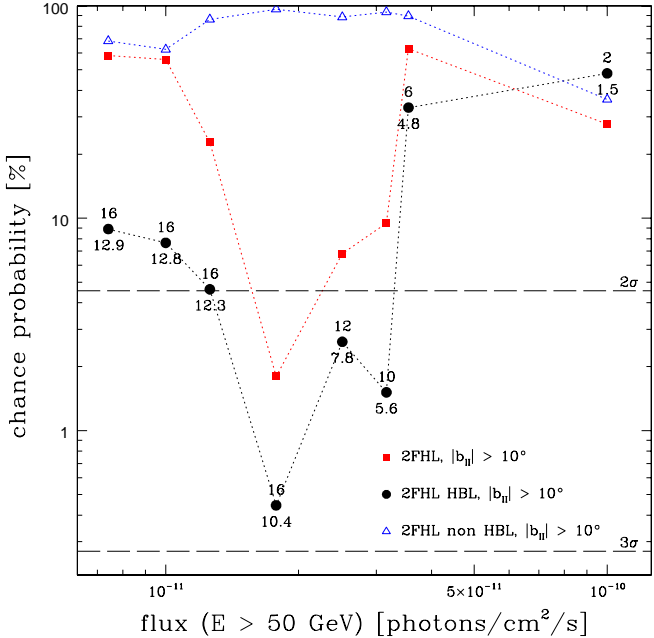}
\caption{The chance probability of association of 2FHL with IceCube events
  for objects having $F(> 50~{\rm GeV})$ larger than the value on the x
  axis for the whole high Galactic latitude sample (red squares), HBL
  (black circles), and non HBL (empty blue triangles). All cases refer to a
  randomisation of the $\gamma$-ray sample on right ascension. The numbers
  give the observed (above the points) and average random values (below the
  points) of $N_{{\nu}}$ for HBL. The two dashed lines denote the 2$\sigma$
  and 3$\sigma$ values.}
\label{fig:2FHL}
\end{figure}

Figure \ref{fig:2FHL} shows the chance probability of association of
2FHL high Galactic latitude sources with IceCube events for objects
having $F(> 50~{\rm GeV})$ larger than the value on the x axis and a
randomisation of the $\gamma$-ray sample on right ascension, which
gives the most conservative result. Red squares indicate the whole
high Galactic latitude sample, black circles denote HBL only, and blue
triangles represent non HBL sources. The numbers give the observed
(above the points) and average random value (below the points) of
$N_{{\nu}}$ for HBL. Figure \ref{fig:2FHL} shows the following:

\begin{itemize}
\item for the whole sample and for HBL the chance probability is strongly
  dependent on $\gamma$-ray flux, with an anti-correlation between
  probability and flux up to $F(> 50~{\rm GeV}) \sim 1.8 \times 10^{-11}$
  photon cm$^{-2}$ s$^{-1}$ with a p-value $< 10^{-4}$ 
  according to a Spearman test. We attribute the turn-over in p-value at
  very large $\gamma$-ray fluxes to small number statistics; 
\item p-values $\sim 1.8$ per cent are reached for $F(> 50~{\rm GeV}) \ga 1.8 \times
  10^{-11}$ photon cm$^{-2}$ s$^{-1}$ for the whole sample;
\item p-values $\sim 0.4$ per cent are reached for $F(> 50~{\rm GeV}) \ga 1.8 \times
  10^{-11}$ photon cm$^{-2}$ s$^{-1}$ for HBL;
\item non HBL sources display no correlation between $\gamma$-ray flux and
  p-value, the latter being always $\ga 40$ per cent. This shows that the
  relatively low p-values reached by the whole sample are driven solely by
  HBL;
\item at the flux at which p-value is minimum for HBL $N_{{\nu}}$ is 16,
  while the average value from the randomisation is 10.4, which means
  that only $\approx 6$ IceCube events have a ``real'' counterpart;
\item for the same flux the number of HBL $\gamma$-ray sources with a
  neutrino counterpart is 27, while the whole ``parent'' $\gamma$-ray
  sample includes 92 sources.
\end{itemize}

At this level of p-value this result would be significant for astronomers
but only at the level of ``a hint'' for physicists, who require a
$5\sigma$ level to claim a ``discovery''. We follow physicists here and
consider this a potentially very interesting result, worth of further
investigations. Therefore, we can say that a hint of an association between the 2FHL high Galactic
latitude HBL and IceCube events is present in the data. 

\subsection{2WHSP}
\begin{figure}
\includegraphics[height=9.0cm]{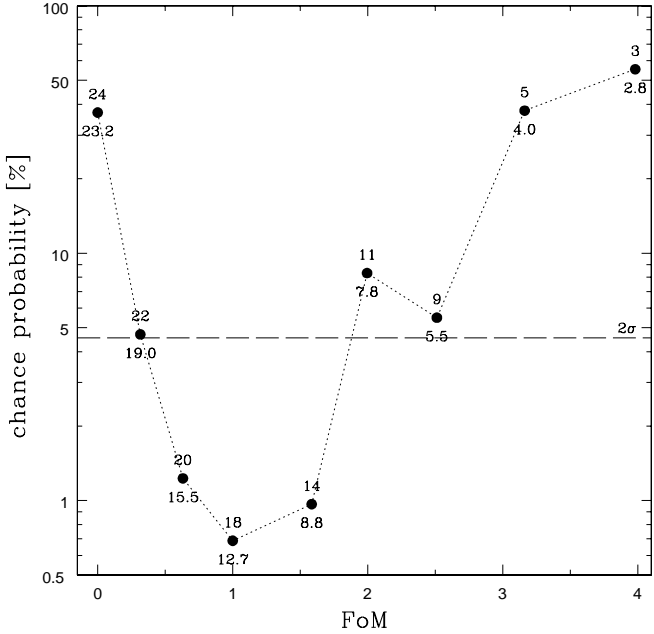}
\caption{The chance probability of association of 2WHSP sources with
  IceCube events for objects having FoM larger than the value on the x axis
  for a randomisation of the $\gamma$-ray sample on both coordinates. The
  numbers give the observed (above the points) and average random values
  (below the points) of $N_{{\nu}}$.  The dashed line denotes the 2$\sigma$
  value.}
\label{fig:2WHSP}
\end{figure}

Figure \ref{fig:2WHSP} shows the chance probability of association of 2WHSP
sources with IceCube events for objects having FoM larger than the value on
the x axis and a randomisation of the $\gamma$-ray sample on both
coordinates, which gives the most conservative result. The numbers give the
observed (above the points) and average random values (below the points) of
$N_{{\nu}}$. The figure shows the following:

\begin{itemize}
\item the chance probability is dependent on FoM, with an anti-correlation
  between probability and flux up to FoM $\sim 1$ with a p-value $<10^{-4}$
  according to a Spearman test;
\item p-values $\sim 0.7$ per cent are reached for FoM $\ga 1$;
\item at the flux at which p-value is minimum for HBL $N_{{\nu}}$ is 18,
  while the average value from the randomisation is 12.7, which means
  that only $\approx 5$ IceCube events have a ``real'' counterpart;
\item for the same flux the number of $\gamma$-ray sources with a neutrino counterpart is 32,
  while the whole ``parent'' $\gamma$-ray sample includes 137 sources.
\end{itemize}

Therefore, a hint of an association between the 2WHSP sample and
IceCube events is present in the data.
Note that a FoM $\sim 1$ is roughly equivalent to $F(> 50~{\rm GeV}) \sim
2.5 \times 10^{-11}$ photon cm$^{-2}$ s$^{-1}$ (de-absorbed). Taking into
account the EBL the corresponding flux is quite close to the value for
which the p-value is minimum for 2FHL HBL.

\subsection{{\it Fermi} 3LAC}

\begin{figure}
\includegraphics[height=9.0cm]{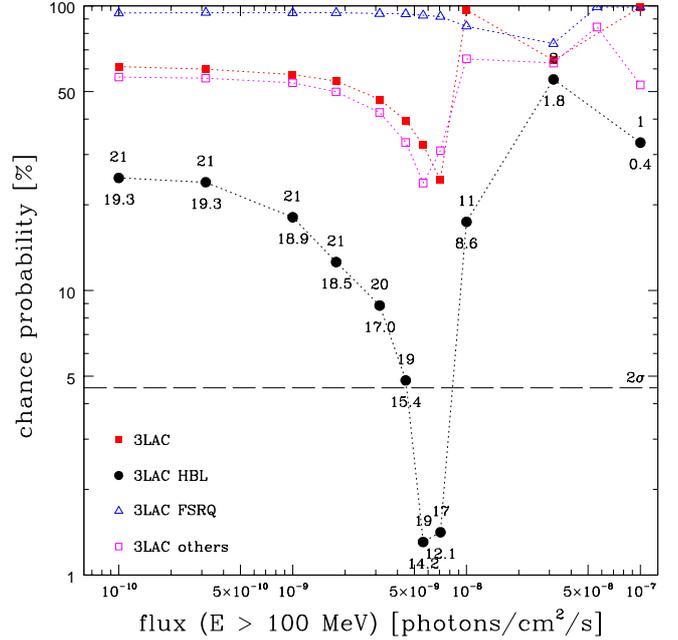}
\caption{The chance probability of association of 3LAC ``clean'' sources
  with IceCube events for objects having $F(> 100~{\rm MeV})$ larger than
  the value on the x axis for the whole sample (filled red squares), HBL
  (black circles), FSRQ (blue triangles), and others (non-HBL and non-FSRQ:
  empty red squares). Only HBL are randomised on right ascension, while the other samples
  are randomised on both coordinates. The
  numbers give the observed (above the points) and average random values
  (below the points) of $N_{{\nu}}$ for HBL. The dashed line denotes the
  2$\sigma$ value.}
\label{fig:3LAC}
\end{figure}

Figure \ref{fig:3LAC} shows the chance probability of association of 3LAC
``clean'' sources with IceCube events for objects having $F(> 100~{\rm
  MeV})$ larger than the value on the x axis. Filled red squares indicate the whole 
sample, randomised on both coordinates, black circles denote HBL, 
randomised on right ascension, blue triangles represent FSRQ, while
empty red squares are for other sources, namely non-HBL and non-FSRQ,
both randomised on both coordinates. 
The numbers give the observed (above the points) and average 
  random values (below the points) of $N_{{\nu}}$ for HBL. 
The figure shows the following:

\begin{itemize}
\item the chance probability is strongly dependent on $\gamma$-ray flux, with 
an anti-correlation between probability and flux up to $F(> 100~{\rm MeV}) \sim 6
  \times 10^{-9}$ photon cm$^{-2}$ s$^{-1}$ with a p-value $< 7 \times 10^{-3}$ 
  for all samples apart from the FSRQ one;
\item the minimum p-value for the whole 3LAC sample and for ``others'' is
  $\sim 24$ per cent for $F(> 100~{\rm MeV}) \sim 6 \times 10^{-9}$ photon
  cm$^{-2}$ s$^{-1}$;
\item the minimum p-value for HBL is $\sim 1.3$ per cent for
$F(> 100~{\rm MeV}) \sim 6 \times 10^{-9}$ photon cm$^{-2}$ s$^{-1}$;
\item the minimum p-value for FSRQ is $\sim 74$ per cent for
$F(> 100~{\rm MeV}) \sim 3 \times 10^{-8}$ photon cm$^{-2}$ s$^{-1}$;
\item at the flux at which the p-value is minimum for HBL
$N_{{\nu}}$ is 19, while the average value from the randomisation is 
14.2, which means that only $\approx 5$ IceCube events have a ``real''
counterpart;
\item for the same flux the number of HBL $\gamma$-ray sources with a neutrino
counterpart is 38, while the whole ``parent'' $\gamma$-ray sample
includes 147 sources.
\end{itemize}

In summary, there is no evidence for an association between the 3LAC
``clean'' sample and IceCube events. As discussed in Section
\ref{sec:def_3LAC}, this was to be expected for the full sample, as
\cite{gluse_2015} did not make any cut on $\gamma$-ray flux. However,
there is a hint of an association between the 3LAC HBL 
and IceCube events. This is not the case for the 3LAC FSRQ. 

\section{Astrophysics of possible IceCube counterparts}

\begin{table*}
\caption{2FHL HBL sources with $F(> 50~{\rm GeV}) \ge 1.8 \times 10^{-11}$
  photon cm$^{-2}$ s$^{-1}$ and 2WHSP sources with FoM $\ge 1.0$ in one
  median angular error radius around the positions of the IceCube
  events. The counterparts of the most probable matches are indicated in
  boldface.}
\begin{tabular}{@{}rllllrrrl}
ID &  2WHSP name &  2FHL name & Common name &  offset & $z$ & FoM & flux$^a$  &Comments \\
          &                    &   & &  deg & & &  & \\
\hline  
 9  &  J091037.0$+$332924 &    J0910.4$+$3327 & Ton 1015 & 11.4 & 0.350 & 2.0 & 0.283 &  positional match (PR14)\\ 
     &   J091552.4$+$293324 &     J0915.9$+$2931 &B2 0912$+$29 &  11.2 &  $>$0.19\magg & 2.5       & 0.324 & positional match (PR14)\\ 
     &  J101504.1$+$492600   &   J1015.0$+$4926 & {\bf 1ES 1011$+$496}  &  15.9 &   0.212  & 4.0 &    1.62\magg  &  most probable match (PR14)\\
     &  J110427.3$+$381231   &     J1104.4$+$3812 & {\bf MKN 421}    &  12.8 & 0.031      & 57.5 &  12.4\magg\magg  & most probable match (PR14)\\
10 &  J235907.8$-$303740    &   & {\bf H 2356$-$309}  & \magg4.7  & 0.165 & 2.0 & 0.69$^b$  & most probable match (PR14)\\
11    & J095302.7$-$084018 &  J0952.9$-$0841 &1RXS J095303.4$-$084003 &   \magg7.0 &  \magg\magg... & 0.8 &  0.385 & positional match (PR14)\\ 
  &  J102243.7$-$011302 &  J1022.7$-$0112 & 1RXS J102244.2$-$011257 & \magg7.7 &  $>$0.36\magg & 1.3 &  0.171 &  positional match (PR14)\\
& J102658.5$-$174858 &  J1027.0$-$1749 &1RXS J102658.5$-$174905  &  \magg9.0 & 0.267 & 1.0 &   0.196 &  most probable match?\\
 12  &  J193656.1$-$471950 & J1936.9$-$4721 &  PMN J1936$-$4719 &  \magg5.6 &  0.265 & 1.3 &   0.240 & most probable match?$^c$\\
      &  J195502.8$-$564028 &  J1954.9$-$5641 &  1RXS J195503.1$-$56403  & \magg4.2 & \magg\magg... & 1.0 & 0.127 & positional match \\                                                       
       &  J195945.6$-$472519 &  J1959.6$-$4725 & SUMSS J195945$-$472519 &  \magg5.9 & \magg\magg... & 1.0  & 0.183 & positional match \\ 
      &   J200925.3$-$484953  &  J2009.4$-$4849 & PKS 2005$-$489 & \magg5.6 & 0.071  &10.0  & 0.970 & positional match (PR14)\\ 
 14  & J171405.4$-$202752 &  J1713.9$-$2027 & 1RXS J171405.2$-$202747 &  \magg9.9 & \magg\magg... & 1.3 & 0.275 & most probable match?\\
 17 &  J155543.0$+$111124 &  J1555.7$+$1111  & {\bf PG 1553$+$113} &  \magg8.9 & \magg\magg... & 7.9  & 4.20\magg & most probable match (PR14)\\
   19  & J050657.8$-$543503 &  J0506.9$-$5434 & 1RXS J050656.8$-$543456 & \magg5.1 & $>$0.26\magg & 1.0 & 0.131 &  positional match (PR14) \\
  &   J054357.2$-$553207  &  J0543.9$-$5533 & 1RXS J054357.3$-$553206  & \magg6.4 &  \magg\magg... & 2.5  & 0.527 & positional match$^d$\\
  20 & J014347.3$-$584551 &  J0143.8$-$5847 &  SUMSS J014347$-$584550 &10.1 & \magg\magg...  & 2.0 & 0.161 &  positional match$^e$\\
   &  J035257.4$-$683117 & J0352.7$-$6831 & PKS 0352$-$686 &  \magg7.6 & 0.087  & 2.0 & 0.228 & positional match (PR14)\\
  22  & J191744.8$-$192131 &  J1917.7$-$1921 &  {\bf 1H1914$-$194} &  \magg4.8 &  0.137 & 1.6 & 0.814 & most probable match (PR14)\\
     & &  J1921.9$-$1607 & PMN J1921$-$1607 & \magg6.7 & \magg\magg... & \magg\magg... &  0.397 &  most probable match?(PR14)$^f$\\  
     & J195814.9$-$301111 & J1958.3$-$3011 & 1RXS J195815.6$-$30111 & \magg9.7  & 0.119 &1.3 &    0.282 & most probable match?(PR14)$^f$\\ 
    26  &  J090534.9$+$135806 &  J0905.7$+$1359 & MG1 J090534$+$1358 & 10.9 & \magg\magg... & 1.0 &  0.192 & positional match (PR14)\\ 
    &  J091552.4$+$293324$^g$   & J0915.9$+$2931 & B2 0912$+$29 &  \magg7.9 &  $>$0.19\magg & 2.5       & 0.324 & positional match (PR14)\\ 
 27 &   J081627.2$-$131152 &  J0816.3$-$1311& PMN J0816$-$1311 &  \magg2.4 &  \magg\magg... & 2.5 &  0.344 & positional match$^e$\\
  35 &  J130421.0$-$435310 &   J1304.5$-$4353 & 1RXS 130421.2$-$435308 & 14.3 & \magg\magg... & 2.0 &   0.235 & positional match (PR14)\\ 
       &  J130737.9$-$425938 &  J1307.6$-$4259 & 1RXS 130737.8$-$425940 & 14.8 &  \magg\magg... & 3.2 & 0.351 & positional match (PR14)\\
       &   J131503.3$-$423649 & J1315.0$-$4238  & 1ES 1312$-$423  & 14.6 & 0.105 & 2.5 &  0.157 & positional match (PR14)\\
       & J132840.6$-$472749 & J1328.6$-$4728 & 1WGA J1328.6$-$4727 & \magg9.2  & \magg\magg... &   0.4 & 0.209 & positional match \\ 
       & J134441.7$-$451007 &  &  SUMSS J134441$-$451002 & 10.7 &  \magg\magg... & 1.0 & & positional match\\                             
    39 &  & J0622.4$-$2604 &  PMN J0622$-$2605 & 12.8 & 0.414  & \magg\magg... &  0.258 & positional match \\ 
         & J063059.5$-$240646 &  J0631.0$-$2406 & 1RXS J063059.7$-$240636 &  10.0 &  \magg\magg... &  1.6 &  0.322  & positional match \\
        &  J064933.6$-$313920 & J0649.6$-$3139 &  1RXS J064933.8$-$31391 & 14.2 &  \magg\magg... &   0.8 &   0.225 & most probable match?\\
  40  &  J102356.1$-$433601 & & SUMSS J102356$-$433600 & \magg9.7 &  \magg\magg... & 2.5 & 2.08$^b$ & most probable match?\\
 41  &  J041652.4$+$010523 &  J0416.9$+$0105 & {\bf 1ES 0414$+$009} & \magg2.9 & 0.287  & 3.2 &  0.269  & most probable match\\
 46   & J094709.5$-$254100 &  & 1RXS J094709.2$-$254056 & \magg4.7 &  \magg\magg... & 1.0  & & positional match\\                                
     & J102658.5$-$174858$^h$ & J1027.0$-$1749 & 1RXS J102658.5$-$174905 &    \magg7.4 & 0.267 &  1.0 &  0.196 &  most probable match?\\
    48  & J144037.8$-$384655 &  J1440.7$-$3847 & 1RXS J144037.4$-$38465 &  \magg8.0 & \magg\magg... &  1.3 & 0.184 & positional match \\
    51 & J054030.0$+$582338 &  J0540.5$+$5822 & GB6 J0540$+$5823 &  \magg4.8 & \magg\magg... & 1.6 & 0.187 & positional match \\
         & J060200.4$+$531600 &  J0601.9$+$5317 & GB6 J0601$+$5315 &   \magg1.3 & \magg\magg... & 1.0  &  0.101 & positional match\\    
\hline
\multicolumn{8}{l}{\footnotesize $^a$ f (E $> 50$ GeV) in units of $10^{-10}$ ph/cm$^2$/s}\\
\multicolumn{8}{l}{\footnotesize $^b$ not in 2FHL: 1FHL flux [f (E $> 10$ GeV)]}\\
\multicolumn{8}{l}{\footnotesize $^c$ was positional match in PR14}\\
\multicolumn{8}{l}{\footnotesize $^d$ was most probable match in PR14 but TeV upper limits rule that out}\\
\multicolumn{8}{l}{\footnotesize $^e$ was most probable match in PR14 but 2FHL data rule that out}\\
\multicolumn{8}{l}{\footnotesize $^f$ was positional match in PR14}\\
\multicolumn{8}{l}{\footnotesize $^g$ also counterpart of ID 9}\\
\multicolumn{8}{l}{\footnotesize $^h$ also counterpart of ID 11}\\
\label{tab:2WHSP_counterparts}
\end{tabular}
\end{table*}

The deviations from random expectation of $N_\nu$ for the 2FHL HBL and the
2WHSP samples are $\sim 0.4$ per cent for $F(> 50~{\rm GeV}) \ga 1.8 \times
10^{-11}$ photon cm$^{-2}$ s$^{-1}$ and $\sim 0.7$ per cent for FoM $\ga 1$
respectively. We therefore examined in detail the corresponding
counterparts to IceCube events, which are all HBL by definition.

Table \ref{tab:2WHSP_counterparts} lists the main properties of the 2FHL
HBL and 2WHSP sources satisfying the requirements mentioned above by giving
the IceCube ID, the 2WHSP name (which includes the coordinates), the 2FHL
name, the common name, the offset between the reconstructed position of the
IceCube event and the blazar one, the redshift of the source (if
available), the FoM, and the $> 50$ GeV flux from the 2FHL catalogue (if
available). Given the very strong variability of blazars the flux values
should be taken only as approximate. Nevertheless, on average a stronger
neutrino source should also be a stronger $\gamma$-ray source, unless
significant absorption is present.

The table contains 37 objects matched to 18 IceCube events.  The
overlap between the two samples, as expected, is quite large: 25 of
the 27 2FHL objects are also 2WHSP sources, although three of them are
below the FoM cut of 1, while 28 of the 32 2WHSP objects are also
2FHL sources, although five of them are below the $\gamma$-ray flux cut
of $F(> 50~{\rm GeV}) \ga 1.8 \times 10^{-11}$ photon cm$^{-2}$
s$^{-1}$. Two sources are matched to two different IceCube events: 
2WHSPJ091552.4$+$293324 (ID 9 and 26) and 2FHL J1027.0$-$1749 (ID 11
and 46).

The comments in Table \ref{tab:2WHSP_counterparts} refer to the hybrid
photon -- neutrino SED of the sources.  Namely, following PR14 we have
first put together the $\gamma$-ray SEDs of all sources using the SED
builder\footnote{http://tools.asdc.asi.it/SED/} of the ASI Science Data
Centre (ASDC) adding, if needed, VHE data taken from the literature. We
have also included the flux per neutrino event at the specific energy. We
then performed an ``energetic'' diagnostic by checking if a simple
extrapolation succeeded in connecting the most energetic $\gamma$-rays to
the IceCube neutrino in the hybrid SED, taking into account the rather
large uncertainty in the flux of the latter. If this was the case we
considered the source to be a ``most probable'' match. Otherwise, the
object was considered a simple positional match. The idea behind this was
to see how the neutrino and photon energetics compared and therefore have a
much stronger discriminant than a simple cross-correlation.

We note that we do not include here the 3LAC counterparts for the
simple reason that the few sources not overlapping with 2FHL or 2WHSP
do not have, by definition, VHE $\gamma$-ray data and therefore the
hybrid SED is not very informative.

As it turns out, the large majority of the sources in Table
\ref{tab:2WHSP_counterparts}, that is all but six of those associated with
the IceCube events discovered in the first three years of data, had already
been considered by PR14. Based on the new data available and the revised
neutrino fluxes (see Section \ref{sec:neutrino_list}), the classifications
made by PR14 have been changed for six sources, as detailed in the Table.

\begin{figure}
\includegraphics[height=9.0cm]{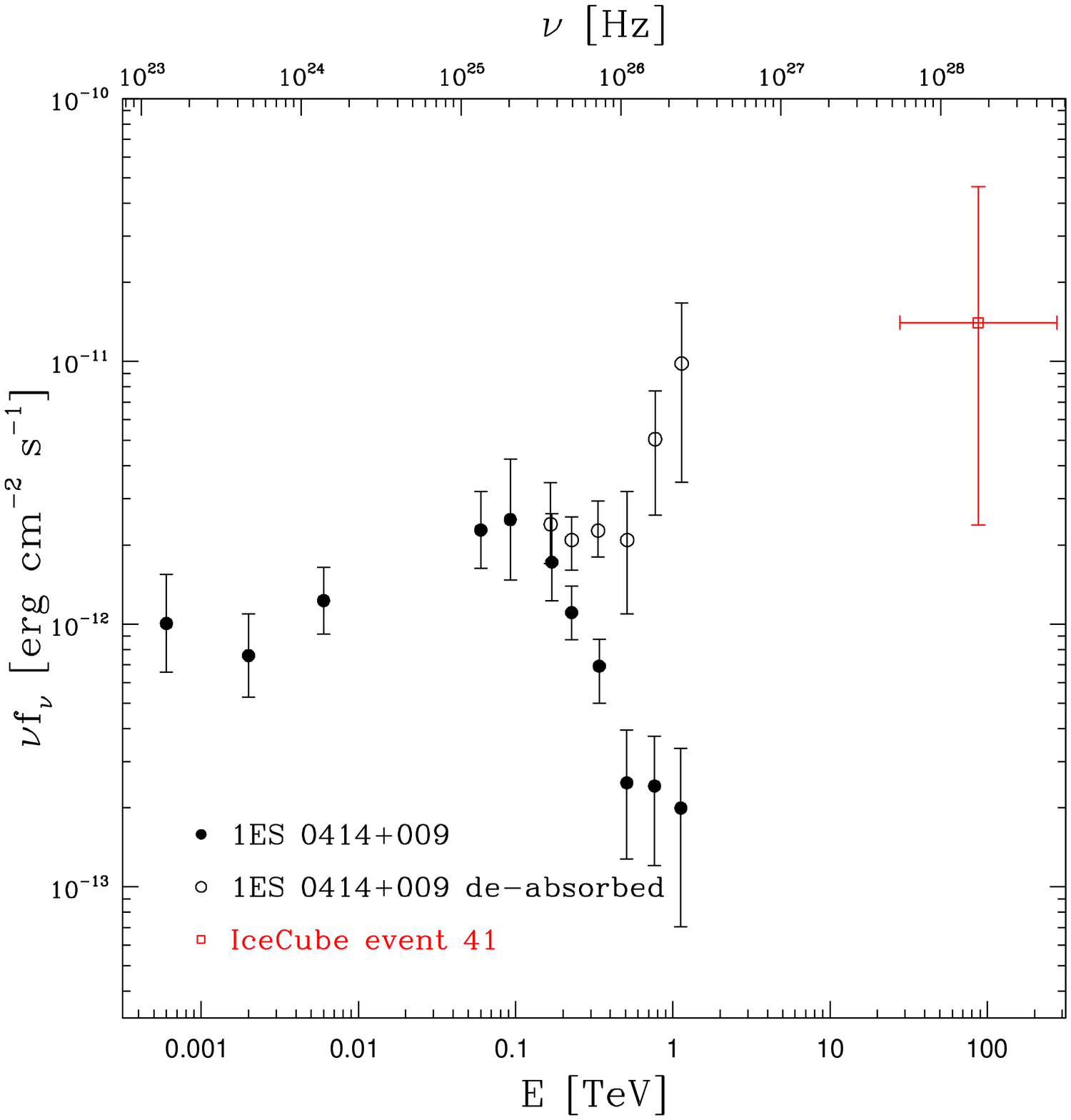}
\caption{{\it Fermi} $\gamma$-ray SED \citep{2FHL,3FGL} and HESS data
  observed (black filled circles) and corrected for absorption by the EBL
  \citep[black open circles;][]{Finke_2015} of 1ES 0414+009, the only 2WHSP
  counterpart with FoM $\ge 2.5$ in the IceCube error circle (ID 41). The
  (red) open square represents the neutrino flux for the corresponding
  IceCube event; vertical error bars are Poissonian for one event, while
  the horizontal one indicates the range over which the flux is
  integrated.}
\label{fig:SED_0414}
\end{figure}

The SEDs of sources associated with IceCube events having ID $> 35$ in
Table \ref{tab:2WHSP_counterparts}, associated with the fourth year of
IceCube data, were studied ex-novo. At least one, that is 1ES
0414+009, whose SED is shown in Fig. \ref{fig:SED_0414}, is without
doubt a most probable match, giving its rising SED and large
de-absorbed TeV fluxes, while three more could be most probable
matches.

The detailed SED study of the 2FHL and WHSP candidates suggests then
that five IceCube events (9, 10, 17, 22, and 41) have most probable
matches (with the respective counterparts highlighted in boldface),
with a few more (11, 12, 14, 39, 40, and 46) having possible
counterparts.  SUMSS J102356$-$433600, however, the possible
counterpart of ID 40, is an unlikely counterpart once one considers
the Galactic source Vela Junior (see Appendix \ref{sec:Galactic} and
Fig. \ref{fig:SED_0852}).

\section{Discussion}


Consistently with PR14, HBL appear to be the only plausible blazar
counterparts of at least some of the IceCube events. However, while in PR14
this statement could not be properly supported by a statistical analysis
due to the lack of complete, all-sky, VHE catalogues, in this paper we have
addressed this issue by introducing a new observable, $N_{{\nu}}$, and by
making use of the very recently completed, and independently built, 2FHL,
3LAC, and 2WHSP catalogues. The former is defined as the number of neutrino
events with at least one $\gamma$-ray counterpart found within the
individual median angular error.

The chance probability in the individual catalogues,
$P_i(N_{{\nu}}(f_\gamma,i))$, reaches values $0.4 - 1.3$ per cent for HBL and
appears to be strongly dependent on $\gamma$-ray flux and FoM, with a
minimum at $F(> 50~{\rm GeV}) \ga 1.8 \times 10^{-11}$ photon cm$^{-2}$
s$^{-1}$ (2FHL), $F(> 100~{\rm MeV}) \ga 5.6 \times 10^{-9}$ photon
cm$^{-2}$ s$^{-1}$ (3LAC), and FoM $\ga 1$. We then carefully examined the
SEDs of all these sources, applying the ``energetic'' test of PR14. This
way we identified $\approx 5$ HBL by checking if a simple extrapolation
succeeded in connecting the most energetic $\gamma$-rays to the IceCube
neutrino in an hybrid SED.

This number of most probable matches is in very good agreement with
the value coming from our randomisations. Namely, we expect $\sim 10,
13$, and 14 spurious matches on average (for the 2FHL, 2WHSP, and 3LAC
catalogues respectively), which translates, based on the number of
observed matches (16, 18, and 19), into $\sim 5 - 6$ ``real''
counterparts.  This highlights once more the importance of the SED
diagnostic in singling out the best candidates.
 

As found by \cite{Pad_2015} in a model-dependent fashion, it turns out
that the neutrino signal from blazars is not a predominant component
of current IceCube data: only $\approx 5$ (up to $\approx 10$ including
the possible ``most probable matches'') IceCube events can be
associated with HBL. These need to be compared with our list of 51
events, which includes 30 HESE and 21 through-going $\nu_{\mu}$, which
corresponds to a model-independent fraction of the IceCube signal $\sim 10 - 20$ per cent. 
This does not exclude the possibility that this fraction might increase
once IceCube reaches fainter fluxes. 


Based on our results, to be a neutrino source candidate a blazar needs
to be: 1. a relatively strong source; 2. a VHE $\gamma$-ray source;
3. an HBL. This is relevant also to explain the lack of signal from
{\it Fermi} blazars in IceCube \cite{gluse_2015}, who, we note, did
not make any cut on $\gamma$-ray flux, although they considered, apart from the whole 2LAC, 
also FSRQ, LBL, and IBL+HBL sub-samples.

We have in fact applied the same statistical tests not only to HBL but
also to FSRQ and IBL and LBL in general, with null results (see
Fig. \ref{fig:2FHL} and \ref{fig:3LAC}). This leaves HBL as
the only possible extragalactic $\gamma$-ray detected IceCube
counterparts. We note that, since PR14 used TeVCat, 1WHSP, and 1FHL, 
the sensitivity to FSRQ was only marginal (e.g. only $\sim 10$ per cent of the sources
in their Table 2 are FSRQ). This is not the case in this paper, since we also 
consider the 3LAC sample. 

None of the matches in Tab. \ref{tab:2WHSP_counterparts} are track-like
events. To probe this further, we have re-done the statistical analysis separately for the 29
track-like events, with null results. Namely, no counterparts were observed for the
full 2FHL and 3LAC HBL samples, while three were detected in the full 2WHSP catalogue (all with
FoM $\le 0.3$). The corresponding curves in Fig. 1 to 3 would then be an horizontal line at 100\% 
for the first two samples and a very similar line (with a small dip to $\sim 20\%$ only for FoM $\le 0.3$) 
for 2WHSP. The lack of track-like events in Tab. \ref{tab:2WHSP_counterparts} persists
even if we assume a median angular error of
$1^{\circ}$ for the $\nu_{\mu}$ events. Only for an error of $2^{\circ}$
one counterpart with FoM $\ge 2$ appears in the 2WHSP catalogue, while
$\sim 0.5$ are expected, based on our simulations. This indicates that by
using tracks only we are still not sensitive to the HBL neutrino signal,
as also expected from the fact that tracks trace only about 1/6 of the 
astrophysical signal under the assumption of a flavour ratio $\nu_{\rm e}:\nu_\mu:\nu_\tau = 1:1:1$. 
We note that \cite{ICECube15_4} have looked for correlations between IceCube 
neutrinos and the highest-energy cosmic rays measured by the Pierre Auger 
Observatory and the Telescope Array. Even in their case the smallest of the p-values 
comes from the correlation between ultrahigh-energy cosmic rays with IceCube 
cascades (i.e. non track-like).

\begin{figure}
\includegraphics[height=9.0cm]{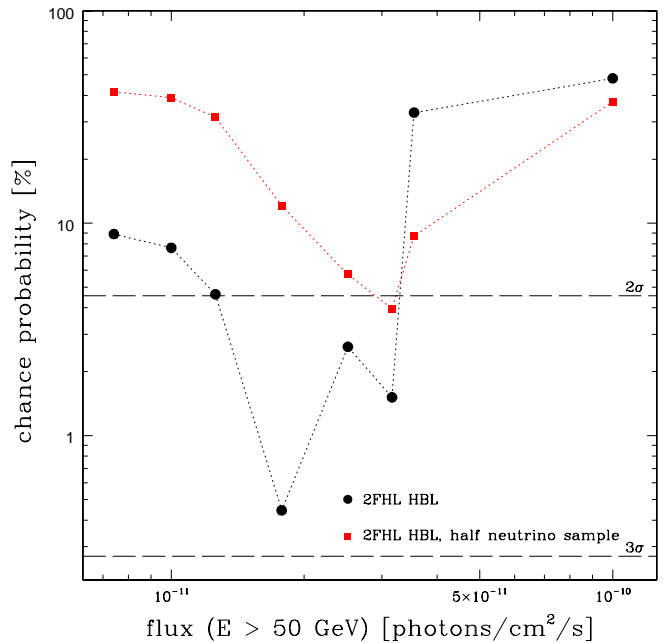}
\caption{The chance probability of association of high Galactic latitude
  2FHL HBL for objects having $F(> 50~{\rm GeV})$ larger than the value on
  the x axis for the four year (black circles) and for the first half of the
  IceCube HESE sample (red squares).  All cases refer to a randomisation of
  the $\gamma$-ray sample on right ascension.}
\label{fig:half_sample}
\end{figure}

It is important to stress that we are not limited by the $\gamma$-ray
samples but by the neutrino statistics. As illustrated, in fact, in
Fig. \ref{fig:half_sample}, comparing the first half of the HESE sample
with the four year one, the p-value decreases steadily at relevant
$\gamma$-ray fluxes for the 2FHL HBL as the live time increases.  Assuming
there is indeed a signal, this gives us hope that the continuous
accumulation of data from IceCube and future neutrino observatories
\citep[e.g.  KM3NeT, IceCube-Gen2:][]
      {2014NIMPA.766...83M,2014arXiv1412.5106I} can turn the hint we
      observed into a discovery.


\section{Conclusions}

We have investigated the correlation between $\gamma$-ray sources from the
2FHL, 2WHSP, and 3LAC samples with the latest list of IceCube neutrinos.
This was done by first deriving the number of neutrino events with at least
one $\gamma$-ray counterpart within the individual IceCube median angular
error and then by estimating the related chance probability using an ensemble
of $10^5 - 10^6$ random maps. For the three catalogues the p-values reach
$0.4 - 1.3$ per cent for HBL and appears to be strongly dependent
on $\gamma$-ray flux and FoM. Through careful examination of the hybrid
$\gamma$-ray -- neutrino SEDs of the sources giving the strongest signal (the
``energetic'' test of PR14) we have identified $\approx 5$ HBL as the most
probable IceCube counterparts. This number is in very good agreement with the
value coming from our randomisations, highlighting once more the importance
of the SED diagnostic in singling out the best candidates, and corresponds to
a model-independent fraction of the current IceCube signal $\sim 10 -
20$ per cent. Other types of blazars give null results, indicating that to be a
neutrino source candidate a blazar needs to be a relatively strong VHE
$\gamma$-ray source with \nup $> 10^{15}$ Hz. The p-values
obtained for the 2FHL HBL by comparing the first half of the HESE sample with
the four year one indicates that we are limited by the neutrino statistics. If a
signal is indeed there, more data from IceCube and future neutrino
observatories should turn our hint into a discovery. 

As for Galactic sources, although we cannot perform a correlation study similar to
that done for blazars due to the complications related to the randomisation
in this case, we nevertheless studied their hybrid SEDs and found that two
IceCube neutrinos have most probable Galactic 2FHL counterparts (with one
more having a possible counterpart: see Appendix A). 

\section*{Acknowledgments}

We thank Stefan Coenders for useful comments and the many teams, which have
produced the data and catalogues used in this paper for making this work
possible. ER is supported by a Heisenberg Professorship of the Deutsche
Forschungsgemeinschaft (DFG RE 2262/4-1), BA by the Brazilian Scientific Program 
``Ci\^encias sem Fronteiras" Cnpq, and YLC by a scholarship provided by 
the Government of the Republic of China (Taiwan). We acknowledge the use of data
and software facilities from the ASDC, managed by the Italian Space Agency
(ASI).


\appendix
\section{Galactic sources}\label{sec:Galactic}

\begin{table*}
\caption{2FHL Galactic sources in one median angular error radius
  around the positions of the IceCube events. The counterparts
of the most probable matches are indicated in boldface.}
\begin{tabular}{@{}rlllrrrll}
ID &  2FHL name &  Common name &  RA (2000) & Dec (2000) & offset & f (E $> 50$ GeV)  & Class$^a$ & Comments \\
          &                &                            &                              &                             &  deg & $10^{-10}$ ph/cm$^2$/s & \\\hline
 14   &      J1745.7$-$2900   & SgrA* &17 45 42.4 &$-$29  00 37 &  1.3&    1.070 &  Ext. Gal. & positional match (PR14)\\ 
        &      J1801.7$-$2358   & HESS J1800$-$240B &18 01 47.0 &$-$23 58 39 &  5.9&  0.569 &  SNR  &most probable match?\\ 
        &      J1801.3$-$2326   & W28 &18  01 21.5 &$-$23 26 24 &  6.2&    1.280 &  SNR  & positional match (PR14)\\ 
        &      J1805.6$-$2136   & W30 &18  05 38.4 &$-$21 36 36 &  8.2&    2.680 &  Ext. Gal. & positional match (PR14)\\ 
 33   &      J1911.0$+$0905  & W49B &19 11 01.4 &  $+$09 05 13 &  4.9&    0.462 &  SNR & positional match (PR14)\\ 
        &      J1923.2$+$1408  & W51 &19 23 16.8 & $+$14 08 24 &  6.6&    0.897 &  SNR  & positional match (PR14)\\ 
 35   &      J1303.4$-$6312    & HESS J1303$-$631&13 02 59.9 &$-$63 12  00 &  9.8&    0.910 &  PWN & positional match (PR14)\\ 
  &      J1355.1$-$6420   & HESS J1356$-$645&13 55 07.1 &$-$64 20 24 &  8.5& 0.833 & PWN & positional match (PR14)\\ 
       &      J1443.2$-$6221    & RCW86 &14 43 16.8 &$-$62 21 00 &  9.1&    0.683 &  SNR  & positional match (PR14)\\ 
       &      J1419.3$-$6047    & HESS J1420$-$607 &14 19 19.1 &$-$60 48 00 &  6.0&    1.650 &  PWN & positional match (PR14)\\ 
       &      J1514.0$-$5915    & HESS J1514$-$591 &15 14 02.3 &$-$59 15 36 & 11.3&    1.490 &  PWN &  positional match (PR14)\\ 
40   &      J0833.1$-$4511    & Vela Pulsar &08 33  09.5 &$-$45 11 24 & 11.2&  1.030 &  PWN & positional match\\ 
       &      J0835.3$-$4511        & PSR J0835-4510 & 08 35 23.7 & $-$45 11 09    & 10.8 &   0.274 &  Radio Pulsar & positional match\\       
       &      J0852.8$-$4631    & {\bf Vela Junior}  &08 52 48.0 &$-$46 31 12 &  7.5&    5.030 &  SNR  & most probable match\\ 
52   &      J1615.3$-$5146    & {\bf HESS J1614$-$518} &16 15 19.2 &$-$51 46 48 &  5.8&    2.340 &  Ext. Gal.  & most probable match\\ 
       &      J1616.2$-$5054    & HESS J1616$-$508 &16 16 14.4 &$-$50 54 36 &  6.2&    1.860 &  PWN & most probable match?\\ 
       &      J1633.5$-$4746    & HESS J1632$-$478&16 33 00.0 &$-$47 46 12 &  6.9&    2.580 &  PWN & most probable match?\\ 
       &      J1640.6$-$4632    & HESS J1641$-$463 &16 40 41.7 &$-$46 33 00 &  7.6&    1.030 &  SNR & positional match \\ 
\hline
\multicolumn{9}{l}{\footnotesize $^a$ SNR: supernova remnant; PWN: pulsar wind nebula; ext. Gal.: extended Galactic source}\\
\label{tab:GAL_counterparts}
\end{tabular}
\end{table*}

The case for Galactic sources is more complex: a randomisation of the
$\gamma$-ray positions is, in fact, in this case meaningless because
one loses the information on the concentration of sources at small
$b_{\rm II}$ ($\la 2^{\circ}$) values. And, as discussed in
Sec. \ref{sec:stat}, the randomisation of the IceCube right ascensions
would result in a biased test statistics.

Nevertheless, based on the results of PR14, who had singled out two
PWN as most probable counterparts to two IceCube neutrinos, we list in
Table \ref{tab:GAL_counterparts} the main properties of all Galactic
2FHL matches with $|b_{\rm II}| \le 10^{\circ}$, defined as non blazar
sources excluding the unclassified ones. The table gives the IceCube
ID, the 2FHL name, the common name, the 2FHL coordinates, the offset
between the reconstructed position of the IceCube event and the blazar
one, the 2FHL $> 50$ GeV flux, and the class. Note that 3/5 events are
also listed in Tab. \ref{tab:2WHSP_counterparts}.

\begin{figure}
\includegraphics[height=9.0cm]{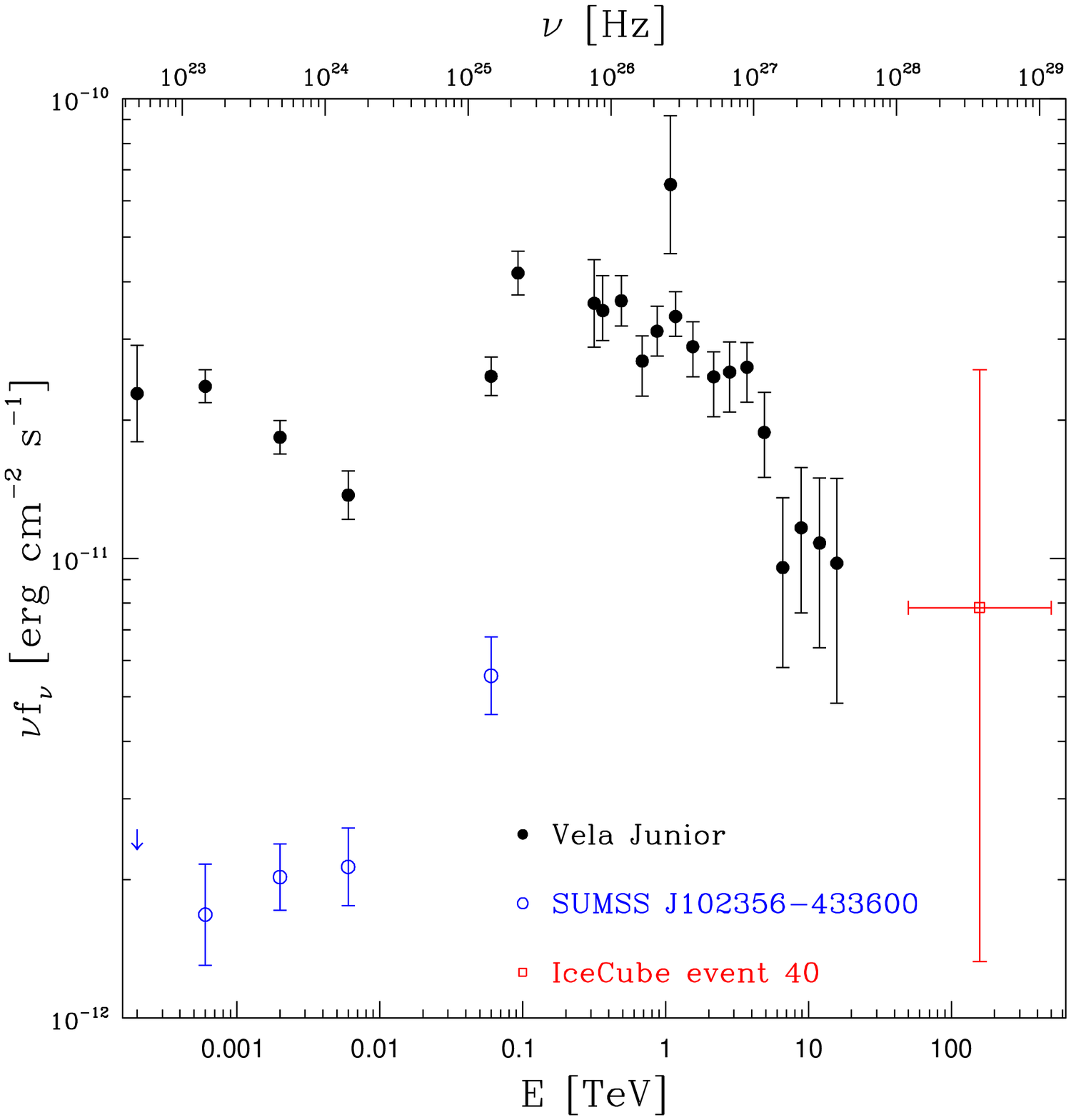}
\caption{$\gamma$-ray SEDs of two sources in the error circle of ID
  40, namely: the Vela Junior SNR \citep[black filled
    circles:][]{2FHL,3FGL,Aharonian_2007} and the HBL SUMSS
  J102356$-$433600 \citep[open blue circles:][]{3FGL}. The (red) open
  square represents the neutrino flux for the corresponding IceCube
  event; error bars as described in the caption of
  Fig. \ref{fig:SED_0414}.}
\label{fig:SED_0852}
\end{figure}

We note that all sources in Table
\ref{tab:GAL_counterparts} associated with the IceCube
events discovered in the first three years of data had been already considered by
PR14. Based on the data available now and the revised neutrino fluxes (see
Section \ref{sec:neutrino_list}), we confirm the classification made by
PR14 for all but one source: HESS J1800$-$240B was considered a positional match
but we now believe it {\it could} be a most probable match. We note that none of the two PWN 
in Table 4 of PR14, namely HESS J1809$-$193 (connected to IceCube event 14) and MGRO 
J1980$+$06 (related to IceCube event 33) are in Table \ref{tab:GAL_counterparts}. This could be
due to their relatively steep $\gamma$-ray spectra. Based on Figs. 6 and 7 of PR14 we still consider
these two sources to be most probable matches. 

\begin{figure}
\includegraphics[height=9.0cm]{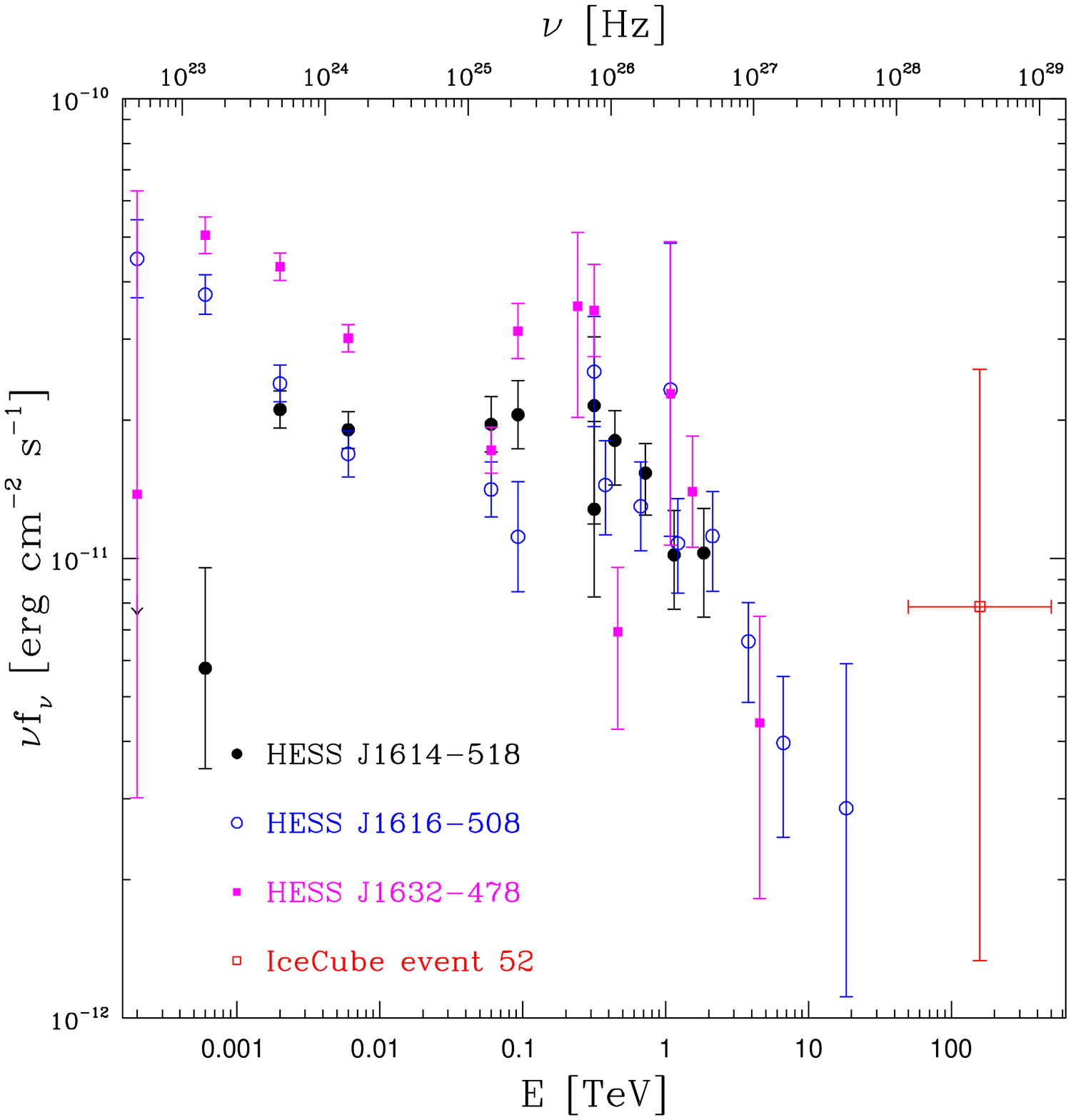}
\caption{$\gamma$-ray SEDs of three sources in the error circle of ID
  52, namely: the extended Galactic source HESS J1614$-$518
  \citep[black filled circles:][]{2FHL,3FGL,Acero_2013}, the PWN HESS
  J1616$-$508 \citep[blue open circles:][]{2FHL,3FGL,Aharonian_2006},
  and the PWN HESS J1632$-$478 \citep[magenta filled
    squares:][]{2FHL,3FGL,Aharonian_2006}. The (red) open square
  represents the neutrino flux for the corresponding IceCube event;
  error bars as described in the caption of Fig. \ref{fig:SED_0414}.}
\label{fig:SED_ID52}
\end{figure}

The SEDs of the last seven 2FHL sources in Table \ref{tab:GAL_counterparts},
associated with the fourth year of IceCube data, were studied ex-novo. Of
the first three, associated with ID 40, Vela Junior is without doubt a most probable
match. Its SED is shown in Fig. \ref{fig:SED_0852}, together with that of
the HBL SUMSS J102356$-$433600. It turns out that the Galactic source is a better candidate. 
Of the last four 2FHL
sources, all associated with ID 52, only the last one is a simple
positional match, while none of the other three can be dismissed on the
basis of the ``energetic'' diagnostic. Their SEDs are shown in
Fig. \ref{fig:SED_ID52}. HESS J1614$-$518 might be more favoured simply
because its SED does not drop at high energies like the other two sources
for lack of data.

The detailed SED study of the 2FHL candidates suggests then that two IceCube
events (40 and 52) have most probable Galactic 2FHL counterparts (with the respective
counterparts highlighted in boldface), with one more (14)
having a possible counterpart.

\label{lastpage}

\bsp	


\begin{thebibliography}{}
\bibitem[\protect\citeauthoryear{Aartsen et
    al.}{2013}]{2013PhRvL.111b1103A} Aartsen M.~G., et al., 2013,
  Phys. Rev. Let., 111, 021103
\bibitem[\protect\citeauthoryear{Aartsen et al.}{2015}]{Aartsen2015}
  Aartsen M.~G., et al., 2015, Phys. Rev. Let., 115, 081102
\bibitem[\protect\citeauthoryear{Acero et al.}{2013}]{Acero_2013} Acero F.,
  et al., 2013, ApJ, 773, 77
\bibitem[\protect\citeauthoryear{Acero et al.}{2015}]{3FGL} Acero F., et
  al., 2015, ApJS, 218, 23
\bibitem[\protect\citeauthoryear{Ackermann et al.}{2011}]{fermi2lac}
  Ackermann M., et al., 2011, ApJ, 743, 171
\bibitem[\protect\citeauthoryear{Ackermann et al.}{2013}]{1FHL} Ackermann
  M., et al., 2013, ApJS, 209, 34
\bibitem[\protect\citeauthoryear{Ackermann et al.}{2015}]{Fermi3LAC}
  Ackermann M., et al., 2015, ApJ, 810, 14
\bibitem[\protect\citeauthoryear{Aharonian}{2004}]{2004vhec.book.....A}
  Aharonian F., 2004, Very High Energy Cosmic Gamma Radiation: A Crucial
  Window on the Extreme Universe. World Scientific Press, Singapore
\bibitem[\protect\citeauthoryear{Aharonian et al.}{2006}]{Aharonian_2006}
  Aharonian F., et al., 2006, ApJ, 636, 777
\bibitem[\protect\citeauthoryear{Aharonian et al.}{2007}]{Aharonian_2007}
  Aharonian F., et al., 2007, ApJ, 661, 236
\bibitem[\protect\citeauthoryear{Ahlers \& Halzen}{2015}]{Ahlers_2015}
  Ahlers M., Halzen F., 2015, Rep. Prog. Phys., 78, 126901
\bibitem[\protect\citeauthoryear{Arsioli et al.}{2015}]{Arsioli2015}
  Arsioli B., Fraga B., Giommi P., Padovani P., Marrese, M., 2015, A\&A,
  579, A34
\bibitem[\protect\citeauthoryear{Brown, Adams, \& Chadwick}{2015}]{brown15}
  Brown A.~M., Adams J., Chadwick P.~M., 2015, MNRAS, 451, 323
\bibitem[\protect\citeauthoryear{The Fermi-LAT Collaboration}{2015}]{2FHL}
  The Fermi-LAT Collaboration, 2015, ApJ, submitted (arXiv:1508.04449)
\bibitem[\protect\citeauthoryear{Finke et al.}{2015}]{Finke_2015} 
Finke J.~D., Reyes L.~C., Georganopoulos M., Reynolds K., Ajello M., Fegan 
S.~J., McCann K., 2015, ApJ, 814, 20   
\bibitem[\protect\citeauthoryear{Gehrels}{1986}]{geh86} Gehrels N.\ 1986,
  ApJ, 303, 336
\bibitem[\protect\citeauthoryear{Giommi et al.}{2012a}]{paper1} Giommi P.,
  Padovani P., Polenta G., Turriziani S., D'Elia V., Piranomonte S., 2012a,
  MNRAS, 420, 2899
\bibitem[\protect\citeauthoryear{Giommi et al.}{2012b}]{GiommiPlanck} Giommi P.,
  et al., 2012b, A\&A, 514, 160
\bibitem[\protect\citeauthoryear{Giommi, Padovani, \& Polenta}{2013}]{paper2}
  Giommi P., Padovani P., Polenta G., 2013, MNRAS, 431, 1914
\bibitem[\protect\citeauthoryear{Giommi \& Padovani}{2015}]{paper4} Giommi
  P., Padovani P., 2015, MNRAS, 450, 2404
\bibitem[\protect\citeauthoryear{Gl{\"u}senkamp et al.}{2015}]{gluse_2015}
  Gl{\"u}senkamp T., for the IceCube Collaboration, 2015, proceedings of
  the RICAP-14 conference, Noto, Sicily (arXiv:1502.03104)
\bibitem[\protect\citeauthoryear{Halzen \& Zas}{1997}]{halzen97} {Halzen} F.,
  {Zas} E., 1997, ApJ, 488, 669
\bibitem[\protect\citeauthoryear{H.E.S.S.~Collaboration et
    al.}{2014}]{HESS_2014} H.E.S.S.~Collaboration, et al., 2014, A\&A, 564,
  A9
\bibitem[\protect\citeauthoryear{IceCube Collaboration}{2013}]{ICECube13}
  IceCube Collaboration, 2013, Science, 342, 1242856
\bibitem[\protect\citeauthoryear{IceCube Collaboration}{2014}]{ICECube14}
  IceCube Collaboration, 2014, Phys. Rev. Lett., 113, 101101
\bibitem[\protect\citeauthoryear{IceCube Collaboration}{2015a}]{ICECube15_1}
  IceCube Collaboration, 2015a, Contributions to the 34th International
  Cosmic Ray Conference (ICRC 2015), p. 45 (arXiv:1510.05223)
\bibitem[\protect\citeauthoryear{IceCube Collaboration}{2015b}]{ICECube15_2}
  IceCube Collaboration, 2015b, Contributions to the 34th International
  Cosmic Ray Conference (ICRC 2015), p. 37 (arXiv:1510.05223)
\bibitem[\protect\citeauthoryear{IceCube Collaboration}{2015c}]{ICECube15_3}
  IceCube Collaboration, 2015c, Contributions to the 34th International
  Cosmic Ray Conference (ICRC 2015), p. 5 (arXiv:1510.05222)
\bibitem[\protect\citeauthoryear{IceCube Collaboration et 
al.}{2015}]{ICECube15_4} IceCube Collaboration, et al., 2015, submitted 
(arXiv:1511.09408) 
\bibitem[\protect\citeauthoryear{IceCube-Gen2
    Collaboration}{2014}]{2014arXiv1412.5106I} IceCube-Gen2 Collaboration,
  2014, arXiv:1412.5106
\bibitem[\protect\citeauthoryear{Kistler, Stanev, \& Y{\"u}ksel}{2014}]{kis14}
  Kistler M.~D., Stanev T., Y{\"u}ksel H., 2014, Phys. Rev. D, 90, 123006   
\bibitem[\protect\citeauthoryear{Mannheim}{1995}]{mannheim95} {Mannheim} K.,
  1995, Astroparticle Physics, 3, 295
\bibitem[\protect\citeauthoryear{Margiotta}{2014}]{2014NIMPA.766...83M}
  Margiotta A., 2014, NIMPA, 766, 83
\bibitem[\protect\citeauthoryear{M{\"u}cke et al.}{2003}]{mueckeetal03}
  {M{\"u}cke} A. et al., 2003, Astroparticle Physics, 18, 593
\bibitem[\protect\citeauthoryear{Murase, Inoue \&
    Dermer}{2014}]{muraseinouedermer14} {Murase} K., {Inoue} Y., {Dermer}
  C.~D., 2014, Phys. Rev. D, 90, 023007
\bibitem[\protect\citeauthoryear{Padovani \& Giommi}{1995}]{padgio95} Padovani
  P., Giommi P., 1995, ApJ, 444, 567  
\bibitem[\protect\citeauthoryear{Padovani \& Resconi}{2014}]{Pad_2014}
  Padovani P., Resconi E., 2014, MNRAS, 443, 474 (PR14)
\bibitem[\protect\citeauthoryear{Padovani \& Giommi}{2015}]{paper3}
  Padovani P., Giommi P., 2015, MNRAS, 446, L41
\bibitem[\protect\citeauthoryear{Padovani et al.}{2015}]{Pad_2015} Padovani
  P., Petropoulou M., Giommi P., Resconi E., 2015, MNRAS, 452, 1877
\bibitem[\protect\citeauthoryear{Petropoulou et al.}{2015}]{Petro_2015}
  Petropoulou M., Dimitrakoudis S., Padovani P., Mastichiadis A., Resconi
  E., 2015, MNRAS, 448, 2412
\bibitem[\protect\citeauthoryear{Schoenen \& Raedel}{2015}]{2015ATel.7856....1S} 
  Schoenen S., Raedel L., 2015, Astron. Telegram, 7856, 1
\bibitem[\protect\citeauthoryear{Urry 
\& Padovani}{1995}]{UP95} Urry C.~M., Padovani P., 1995, PASP, 107, 803   
\bibitem[\protect\citeauthoryear{Tavecchio \& Ghisellini}{2015}]{tav15}
  Tavecchio F., Ghisellini G., 2015, MNRAS, 451, 1502
\end{thebibliography}
\end{document}